\documentclass[11pt]{article}

\usepackage{amsmath}
\usepackage{amsfonts}
\usepackage{amssymb}
\usepackage{graphicx,color}
\usepackage{subfigure}
\usepackage{hyperref}
\usepackage[usenames,dvipsnames,svgnames,table]{xcolor}
\usepackage{enumerate}

\usepackage[left=2cm,top=2.5cm,right=2cm,bottom=2cm]{geometry}
\def \be  {\begin{equation}}
\def \ee  {\end{equation}}
\def \ba  {\begin{eqnarray}}
\def \ea  {\end{eqnarray}}

\begin{document}

\title{General bounds in Hybrid Natural Inflation  }
\bigskip
\author{ Gabriel Germ\'an$^{a}
\footnote{Corresponding author: gabriel@fis.unam.mx}$\,\,,
Alfredo Herrera-Aguilar$^{b,c}$, Juan Carlos Hidalgo$^{a}$, \\Roberto A. Sussman$^{d}$, Jos\'e Tapia$^{a,e}$
\\
\\
{\normalsize \textit{$^a$Instituto de Ciencias F\'isicas,} }\\
{\normalsize \textit{Universidad Nacional Aut\'onoma de M\'exico,}}\\
{\normalsize \textit{Apdo. Postal 48-3, C.P. 62251 Cuernavaca, Morelos, M\'{e}xico.}}
\\
\\
{\normalsize \textit{$^b$Instituto de F\'{\i}sica,} }\\
{\normalsize \textit{Benem\'erita Universidad Aut\'onoma de Puebla,}}\\
{\normalsize \textit{Apdo. Postal J-48, C.P. 72570 Puebla, Puebla, M\'{e}xico.}}
\\
\\
{\normalsize \textit{$^c$Instituto de F\'{\i}sica y Matem\'{a}ticas,}}\\
{\normalsize \textit{Universidad Michoacana de San Nicol\'as de Hidalgo,}}\\
{\normalsize \textit{Edificio C--3, Ciudad Universitaria, C.P. 58040 Morelia, Michoac\'{a}n, M\'{e}xico.}}
\\
\\
{\normalsize \textit{$^d$Instituto de Ciencias Nucleares,} }\\
{\normalsize \textit{Universidad Nacional Aut\'onoma de M\'exico,}}\\
{\normalsize \textit{Apdo. Postal 70Ð543, 04510 M\'exico D. F., M\'exico.}}
\\
\\
{\normalsize \textit{$^e$Centro de Investigaci\'on en Ciencias,} }\\
 {\normalsize \textit{Universidad Aut\'onoma del Estado de Morelos,}}\\
{\normalsize \textit{Avenida Universidad 1001, Cuernavaca, Morelos 62209, M\'exico.}}
}
\date{}
\maketitle

\begin{abstract}
Recently we have studied in great detail a model of Hybrid Natural Inflation (HNI) by constructing two simple effective field theories. These two versions of the model allow  inflationary energy scales as small as the electroweak scale in one of them or as large as the Grand Unification scale in the other, therefore covering the whole range of possible energy scales. In any case the inflationary sector of the model is of the form $V(\phi)=V_0 \left(1+a \cos(\phi/f)\right)$ where $0\leq a<1$ and the end of inflation is triggered by an independent waterfall field. One interesting characteristic of this model is that the slow-roll parameter $\epsilon(\phi)$ is a non-monotonic function of $\phi$ presenting a {\it maximum} close to the inflection point of the potential. Because the scalar spectrum $\mathcal{P}_s(k)$ of density fluctuations when written in terms of the potential is inversely proportional to $\epsilon(\phi)$ we find that $\mathcal{P}_s(k)$ presents a {\it minimum} at $\phi_{min}$. 
The origin of the HNI potential can be traced to a symmetry breaking phenomenon occurring at some energy scale $f$ which gives rise to  a (massless) Goldstone boson. Non-perturbative physics at some temperature $T<f$ might occur which provides a potential (and a small mass) to the originally massless boson to become the inflaton (a pseudo-Nambu-Goldstone boson). Thus the inflaton energy scale $\Delta$ is bounded by the symmetry breaking scale, $\Delta\equiv V_H^{1/4} <f.$ To have such a well defined origin and hierarchy of scales in inflationary models is not common. We use this property of HNI to determine bounds for the inflationary energy scale $\Delta$ and for the tensor-to-scalar ratio $r$.

\end{abstract}

%%%%%%%%%%%%%%%%%%%%%%%%%%%%%%%%%%%%%%%%%%%
%%%%%%%%%%%%%%%%%%%%%%%%%%%%%%%%%%%%%%%%%%%

\section{Introduction} \label{Intro} 

\noindent 
In a recent article \cite{Ross:2016hyb} a model of inflation   \cite{Guth:1980zm}, \cite{Linde:1981mu}, \cite{Albrecht:1982wi}, \cite{Lyth:1998xn} of the hybrid type  \cite{Linde:1994} has been studied with great detail. To show that it is posible within Hybrid Natural Inflation (HNI) \cite{Ross:2016hyb} to account for inflationary energy scales as small as the electroweak scale, or as large as the Grand Unification scale, two versions of the model have been constructed based on simple effective field theories. The resulting inflationary sector in any case is described by the following potential for the inflaton field $\phi$
\begin{equation}
V(\phi) = V_0\left(1+a\cos \left(\frac{\phi}{f} \right) \right),
\label{pot}
\end{equation}
where $a$ is a positive constant less than one and $f$ is the scale of (Nambu-Goldstone) symmetry breaking. 
Here the end of inflation is triggered by an independent sector waterfall field in a rapid phase transition. The potential in Eq.\,(\ref{pot}) is reminiscent of Natural Inflation \cite{Freese:1990rb}, \cite{Adams:1992bn}, \cite{Freese:2014nla} where $a=1$ sets a vanishing cosmological constant. Here, however, $a$ can take any positive value less than one and as a result the scale $f$ can be sub-Planckian. Once the waterfall field triggers the end of inflation the inflaton fast rolls to a global minimum with vanishing energy.

\begin{figure}[t!]
 \begin{center}
   \includegraphics[ width=8cm, height=6cm]{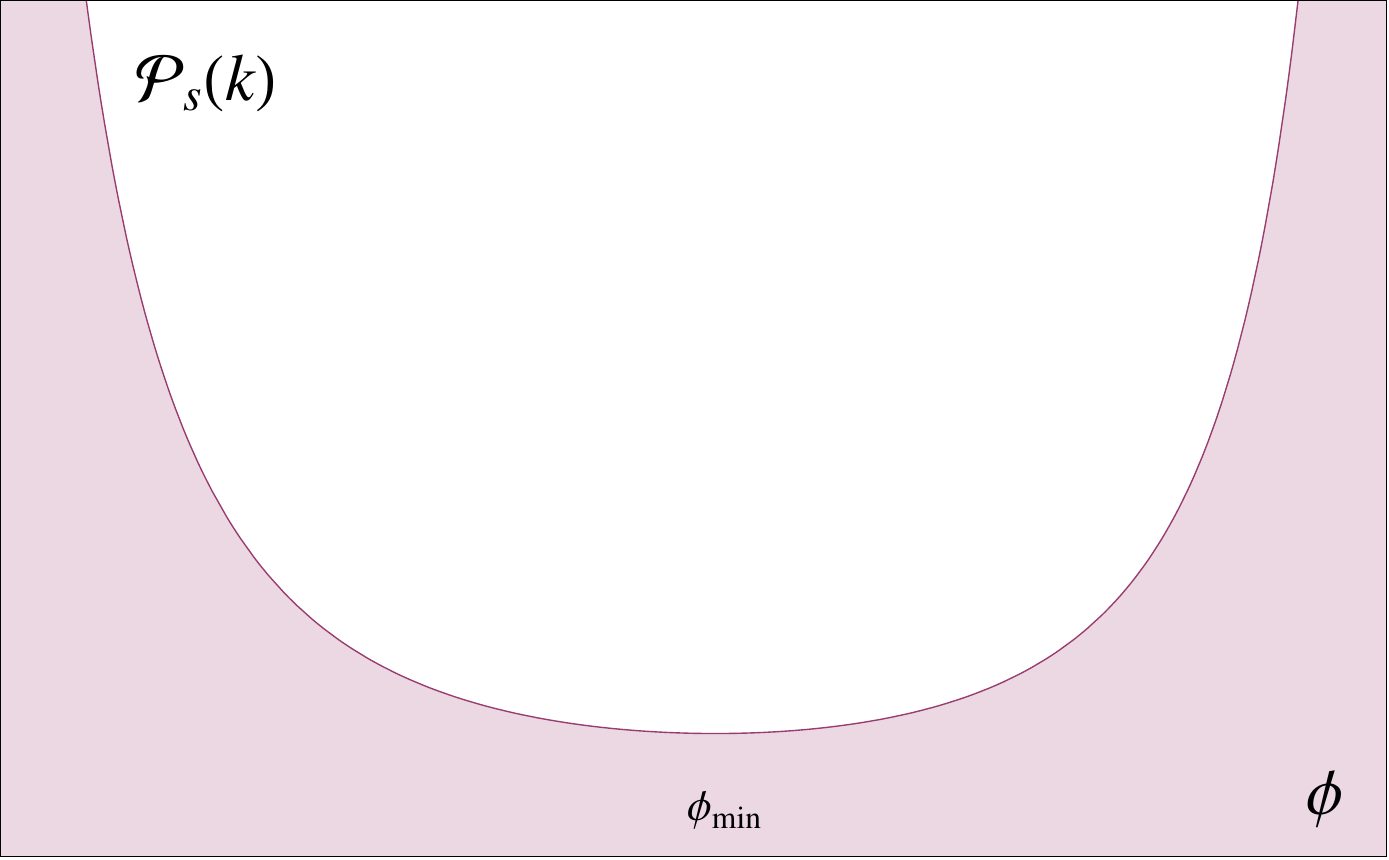}
        \caption{Schematic plot of the scalar density perturbation which can be written in terms of the Hybrid Natural Inflation potential and its derivative as in  Eq.\,(\ref{contraste}). The existence of a {\it maximum} for $\epsilon(\phi)$ is here reflected in the presence of a {\it minimum} for the spectrum. This fact together with the condition that the spectrum decreases during at least 8 e-folds of observable inflation from $\phi_H$ to $\phi_{min}$ is here used to obtain bounds for the scale of inflation $\Delta$ and for the tensor-to-scalar ratio $r$. }
\label{Espectro}
 \end{center}
\end{figure}

Hybrid Natural Inflation has also the interesting property that the slow-roll parameter $\epsilon(\phi)$ turns out to be a non-monotonic function of the inflaton \cite{German:2015qjq}. As a consequence the scalar spectrum of density perturbations develops a {\it minimum} (Fig.\,\ref {Espectro}) for a value $\phi_{min}$ close to the inflection point of the potential.
We know that inflation in HNI should start before $\phi$ reaches the minimum of the spectrum at $\phi_{min}$ because the spectrum has been observed to be  decreasing during at least 8 e-folds of observable inflation. Thus, there should be at least 8 e-folds of inflation from $\phi_H$, at which observable perturbations are produced\footnote{All quantities with a subindex ${}_H$ are evaluated at the scale $\phi_{H}$, at which observable perturbations are produced, some $50-60$ e-folds before the end of inflation.}, to $\phi_{min}$. This minimum amount of inflation with decreasing spectrum should give an upper bound for the tensor-to-scalar ratio $r$ and for the scale of inflation $\Delta$. The remaining $42-52$ e-folds of inflation would occur with an steepening spectrum thus care should be taken to not over-produce primordial black holes (PBH)   \cite{Kohri:2007qn}, \cite{Josan:2009qn}, \cite{Carr:2009jm}. Also the fact that the inflationary energy scale $\Delta$ is bounded by the symmetry breaking scale $f$ imposes {\it lower} bounds to these quantities, whenever the minimum of the spectrum is reached after $N_{min}\leq 60$ e«folds of inflation. If all of inflation occurs without $\phi$ reaching $\phi_{min}$ no lower bounds are found.

Our paper is organised as follows: in Section \ref{slow} we briefly recall expressions for the slow-roll parameters and observables. We also give an effective field theory derivation of the model we study and the hierarchy of energy scales is discussed. As a warming up exercise we initially study in Section \ref{NI}  this hierarchy of scales in Natural Inflation (where $a=1$) and Section \ref{ENI} deals with "extended" Natural Inflation (ENI) where $a$ is not set to unity from the beginning. This allows us to study the fine tuning of $a$ (to have a vanishing cosmological constant) in terms of the parameters of the model. From here we proceed in Section \ref{restricted} to HNI where the hierarchy of energy scales together with the observation that the scalar spectrum is decreasing during  $8<N_{min}<60$ e-folds of observable inflation determine bounds for the inflationary energy scale $\Delta$ and for the tensor-to-scalar ratio $r$, this we call the restricted case. In Section \ref{general} we obtain general bounds in HNI dropping the previous requirement that $N_{min}$ e-folds of inflation occur with decreasing spectrum. We are able to find general bounds for all the parameters (and observables) of the model and to clearly understand how the scale of inflation in HNI is able to sweep all range of values, from vanishingly small to GUT scales. A brief discussion of constraints coming from Primordial Black Hole (PBH) abundances and considerations regarding low scales of inflation can be found in Section \ref{PBH}. Finally Section \ref{conclusions} contains our conclusions and a discussion of the main results.

\section{Slow-roll parameters, observables and model construction } \label{slow}

\noindent
In slow-roll inflation, the spectral indices are given in terms of the slow-roll parameters of the model, which involve the potential $V(\phi)$ and its derivatives (see e.g. \cite{Liddle:94},
\cite{Liddle:2000cg})
\begin{equation}
\epsilon \equiv \frac{M^{2}}{2}\left( \frac{V^{\prime }}{V }\right) ^{2},\quad
\eta \equiv M^{2}\frac{V^{\prime \prime }}{V}, \quad
\xi_2 \equiv M^{4}\frac{V^{\prime }V^{\prime \prime \prime }}{V^{2}},\quad
\xi_3 \equiv M^{6}\frac{V^{\prime 2 }V^{\prime \prime \prime \prime }}{V^{3}},
\label{Slowparameters}
\end{equation}%
primes denote derivatives with respect to the inflaton $\phi$ and $M$ is the
reduced Planck mass $M=2.44\times 10^{18} \,\mathrm{GeV}$. In what follows we set
$M=1$. In the slow-roll approximation observables are given by (see e.g.  \cite{Liddle:2000cg})
\begin{eqnarray}
n_{t} &=&-2\epsilon =-\frac{r}{8} , \label{Int} \\
n_{s} &=&1+2\eta -6\epsilon ,  \label{Ins} \\
n_{sk} &=&\frac{d n_{s}}{d \ln k}=16\epsilon \eta -24\epsilon ^{2}-2\xi_2, \label{Insk} \\
n_{skk} &=&\frac{d^{2} n_{s}}{d \ln k^{2}}=-192\epsilon ^{3}+192\epsilon ^{2}\eta-
32\epsilon \eta^{2} -24\epsilon\xi_2 +2\eta\xi_2 +2\xi_3, \label{Inskk} \\
\mathcal{P}_s(k)&=&\frac{1}{24\pi ^{2}}\frac{V}{\epsilon }=A_s \left( \frac{k}{k_H}\right)^{n_s-1} .
\label{IA} 
\end{eqnarray}
Here  $n_{sk}$ denotes the running of the scalar index and $n_{skk}$ the running of the running, in a self-explanatory notation. All the quantities with a subindex ${}_H$ are evaluated at the scale $\phi_{H}$, at which observable perturbations are produced, some $50-60$ e-folds before the end of inflation. The density perturbation at wave number $k$ is $\mathcal{P}_s(k)$ with amplitude at horizon crossing given by $\mathcal{P}_s(k_H)\approx 2.2\times 10^{-9}$ \cite{Ade:2015xua}, the scale of inflation is $\Delta$ with $\Delta \equiv V_{H}^{1/4}$. The tensor power spectrum parameterised at first order in the SR parameters is
\begin{equation}
\mathcal{P}_t(k)=A_t \left( \frac{k}{k_H}\right)^{n_t} ,
\label{PotHNI}
\end{equation}
it allows to define the tensor-to-scalar ratio as $r\equiv \mathcal{P}_t(k)/\mathcal{P}_s(k)$. 

The construction of the model can proceed by initially considering a potential of the form
\begin{equation}
V\left(\Phi\right) = V_1\left(\Phi\right)+V_2\left(\Phi\right),
\label {V}
\end{equation}
where
\begin{equation}
V_1\left(\Phi\right)= -m^2|\Phi|^2+\lambda |\Phi|^4+\bar{\Delta}^4.
\label {Phi}
\end{equation}
is the potential invariant under the $U(1)$ symmetry, $\Phi \rightarrow e^{i\alpha}\Phi$. The origin of the constant term $\bar{\Delta}^4$ above can be traced to terms in the higher energy sector of the theory. For positive mass-square $m^2$, $\Phi$ triggers spontaneous breaking of the $U(1)$ symmetry and the field $\Phi$ gets a vev given by $\tilde{f}$
\begin{equation}
<\Phi_0>\, = \tilde{f}= \frac{m}{\sqrt{2\lambda}}.
\label {Phi}
\end{equation}
Thus, the potential can be better parameterised by
\begin{equation}
\Phi= \frac{1}{\sqrt{2}}\left(\rho+\tilde{f}\right)\, e^{i\frac{\phi}{\tilde{f}}},
\label {Phi}
\end{equation}
where $\rho$ is a radial field around the minimum of the potential and $\phi$ a Goldstone boson associated with the $U(1)$ symmetry breaking.

The term $V_2\left(\Phi\right)$ in Eq. (\ref{V}) explicitly breaks the $U(1)$ symmetry and generates a mass for the Goldstone boson becoming $\phi$ a Pseudo Nambu-Goldstone boson. We can write a very simple form for $V_2$
\begin{equation}
V_2\left(\Phi\right)=\mu^2\left(\Phi^2+ \Phi^{*2}\right) \sim  \frac{1}{2}\mu^2\tilde{f}^2\left(e^{i\frac{2\phi}{\tilde{f}}} + e^{-i\frac{2\phi}{\tilde{f}}}\right) \sim  \mu^2\tilde{f}^2\cos\left(\frac{2\phi}{\tilde{f}} \right)\ ,
\label {explicit}
\end{equation}
where $\mu$ is a constant with mass dimensions. It follows that the axion potential is
\begin{equation}
V(\phi) =V_0\left[1+a\cos\left(\frac{\phi}{f} \right)   \right],
\label {pot1}
\end{equation}
with $V_0$ defined by  $V_0\equiv \bar{\Delta}^4-\frac{m^4}{4\lambda}$, $\tilde{f}=2 f$ and 
\begin{equation}
a\equiv \frac{4\mu^2 f^2}{V_0}. 
\label {apar}
\end{equation}
The parameter $a$ is bounded as $0\leq a\leq1$. The limiting value $a=0$ reduces the potential to a cosmological constant while $a=1$ gives Natural Inflation.

The origin of the two scales occurring in the potential Eq.\,(\ref{pot1}) is, in principle, well understood and we also know that these scales satisfy a hierarchy: a symmetry breaking phenomenon at some energy scale $f$ gives rise to  a (massless) Goldstone boson while non-perturbative physics at temperature $T<f$ provides a potential (and a small mass) to the originally massless boson becoming a pseudo-Nambu-Goldstone boson, the inflaton. 
In the case of the $QCD$ axion, for example, non-perturbative effects are due to instantons.
Thus the inflationary energy scale $\Delta$ is bounded by the symmetry breaking scale $f$ as $\Delta\equiv V_H^{1/4} \approx V_0^{1/4} < f.$ In the following sections we use this hierarchy of scales to extract bounds for the observables.
%%%%%%%%%%%%%%%%%%%%%%%%%%%%%%%%%%%%%%%%%%%
%%%%%%%%%%%%%%%%%%%%%%%%%%%%%%%%%%%%%%%%%%%

\section{Natural Inflation } \label{NI} 

As a warming up exercise we beguin with Natural Inflation (NI) \cite{Freese:1990rb}. The potential for the NI model  is given by fixing $a=1$ in Eq.\,(\ref{pot1}) above
\begin{equation}
V(\phi) = V_0\left(1+\cos \left(\frac{\phi}{f} \right) \right).
\label{Npot}
\end{equation}
The coefficient of the $\cos \left(\frac{\phi}{f} \right)$-term has been fine-tuned to one so that the potential vanishes at its minimum. We will see that in NI the hierarchy $\Delta < f$ arises in a very natural way. Defining $c_{\phi}\equiv \cos \left( \frac{\phi }{f}\right)$ the slow-roll parameters relevant for what follows are 
\begin{eqnarray}
\epsilon &=&\frac{1}{2f^2}\frac{1-c_{\mathrm{\phi}} }{ 1+c_{\mathrm{\phi}} } ,
\label{NIeps}%
\\
\eta &=&-\frac{1}{f^2}\, \frac{c_{\mathrm{\phi}}}{1+c_{\mathrm{\phi}}} , 
\label{NIeta}
\end{eqnarray}%
Defining $\delta_{n_s}\equiv 1-n_{s_H}$, $c_{H}\equiv \cos \left( \frac{\phi _{H}}{f}\right)$ and $c_{e}\equiv \cos \left( \frac{\phi _{e}}{f}\right)$, the functions at the observable scale and at the end of inflation, the equation for the spectral index Eq.\,(\ref{Ins}) at $\phi _H$, can be written as
\begin{equation}
\delta_{n_s}= \frac{3 -c_H}{f^2(1+c_{H})},
\label{NIspectral}
\end{equation}
solving for $c_H$ we get
\begin{equation}
c_{H_{NI}}\equiv\frac{3-f^2\delta_{n_s}}   {1+f^2\delta_{n_s}},
\label{Nch}
\end{equation}
where the subindices $NI$ means that $c_H$ has been determined from the NI potential Eq.\,(\ref{Npot}). In the following sections dealing with the Hybrid Natural Inflation (HNI) potential instead of writing $HNI$ subindices, they will be simply dropped. 
The end of inflation is given by the saturation of the condition $\epsilon = 1$
\begin{equation}
c_{e_{NI}}\equiv\frac{1-2f^2}   {1+2f^2}.
\label{Nce}
\end{equation}
The number of e-folds from $\phi_H$ to the end of inflation at $\phi _e$ is given by
\begin{equation}
N\equiv -\int_{\phi _H}^{\phi_e}\frac{V({\phi })}{V^{\prime }({\phi })}{d}{\phi }=f^2\ln \left( \frac{1-c_{e_{NI}}}{1-c_{H_{NI}}}\right)=f^2\ln\left[\frac{2f^2(1+f^2 \delta_{n_s})}{(1+2f^2) (-1+f^2 \delta_{n_s}) }  \right],
\label{NN}
\end{equation}
from where it follows that in NI consistency demands that
\begin{equation}
f > \frac{1}{\sqrt{\delta_{n_s}}}.
\label{Nfbound}
\end{equation}
Assuming the spectral index $n_s$ is known, the number of e-folds $N$ fixes the parameter $f$ through Eq.\,(\ref{NN}) which then fixes $\phi_H$ and $\Delta$. For numerical results we take $n_s=0.965$ \cite{Ade:2015xua} thus $\delta_{n_s}=0.035$ and $f \gtrsim 5.3$. 
From Eq.\,(\ref{IA}) the scale of inflation can also be expressed in terms of $f$
\begin{equation}
\Delta=\left(24\pi ^{2}\epsilon(\phi_H)\mathcal{P}_s(k_H)\right)^{1/4} =\left(12\pi^2 \mathcal{P}_s(k_H) \frac{1-c_{H_{NI}}  }{f^2\left(1+\,c_{H_{NI}}\right)}\right)^{1/4}=\left(6\,\pi^2 \delta_{n_s}\mathcal{P}_s(k_H)\left(1-\frac{1}{f^2 \delta_{n_s}}\right) \right)^{1/4},
\label{NDelta}
\end{equation}
from where it follows
\begin{equation}
\Delta < \left(6\pi^2\, \delta_{n_s}\mathcal{P}_s(k_H)\right)^{1/4},
\label{NDeltabound}
\end{equation}
or $\Delta \lesssim 8.3\times 10^{-3} M\approx 2\times 10^{16}\, GeV$.  From Eqs.\,(\ref{Nfbound}) and (\ref{NDeltabound}) we find that 
\begin{equation}
\Delta < \left(6\pi^2\delta_{n_s}\mathcal{P}_s(k_H)\right)^{1/4}< \left(6\pi^2\delta_{n_s}^3\mathcal{P}_s(k_H)\right)^{1/4} f\approx 1.6 \times 10^{-3} f < f,
\label{NDeltalessf}
\end{equation}
thus $\Delta < f$ always. We will see that for this to be the case in HNI the parameter $a$ in the potential has to be bounded from above. The resulting bound for $r$ coming from Eqs.\,(\ref{IA}) and (\ref{NDeltabound}) is
\begin{equation}
r= \frac{2 \Delta^4}{3 \pi^2 \mathcal{P}_s(k_H)} < 4\delta_{n_s}\approx 0.14.
\label{NDeltalessf}
\end{equation}
Of course, in NI, once we have determined $f$ by fixing the number of e-folds $r$ follows: for $\delta_{n_s}=0.035$, Eq.\,(\ref{NN}) with $N=60$  requires $f\approx 8.45$  giving a value $r\approx 0.084$. From Eqs.\,(\ref{NDelta}) and Eq.\,(\ref{apar}) we find that tuning $a=1$ is equivalent to tuning the $\mu$ parameter to the value
\begin{equation}
\mu^2=\frac{3\,\pi^2}{2 f^2} \left(1-\frac{1}{f^2 \delta_{n_s}}\right) \delta_{n_s}\mathcal{P}_s(k_H),
\label{lambda2}
\end{equation}
using the numerical values above, Eq.\,(\ref{lambda2}) gives $\mu\approx 3.1 \times 10^{-6} M\approx 7.6 \times 10^{12}GeV$.
%%%%%%%%%%%%%%%%%%%%%%%%%%%%%%%%%%%%%%%%%%%
%%%%%%%%%%%%%%%%%%%%%%%%%%%%%%%%%%%%%%%%%%%

\section{Extended Natural Inflation } \label{ENI} 

We now make a simple extension of the NI model discussed above leaving the parameter $a$ as coefficient of the $\cos \left(\frac{\phi}{f} \right)$-term in the potential Eq.\,(\ref{pot1}) . This parameter is now restricted to the interval $0<a<1$. The potential is thus
\begin{equation}
V(\phi) = V_0\left(1+a \cos \left(\frac{\phi}{f} \right) \right).
\label{Hpot}
\end{equation}
A potential like Eq.\,(\ref{Hpot}) has been studied in the context of Hybrid Natural Inflation (HNI)  \cite{Ross:2016hyb}, \cite{Ross:2009hg}, \cite{Ross:2010fg}, where the symmetry breaking scale is sub-Planckian and the end of inflation is triggered by an additional waterfall field. Here we would like to study a simple extension of NI (ENI) allowing (as in NI) super-Planckian values for $f$. This would allow us to study the fine tuning of $a$ in terms of the parameters of the model and see what requirements (if any) the inflationary dynamics impose on them.
 
\begin{figure}[t!]
 \begin{center}
   \includegraphics[ width=10cm, height=6cm]{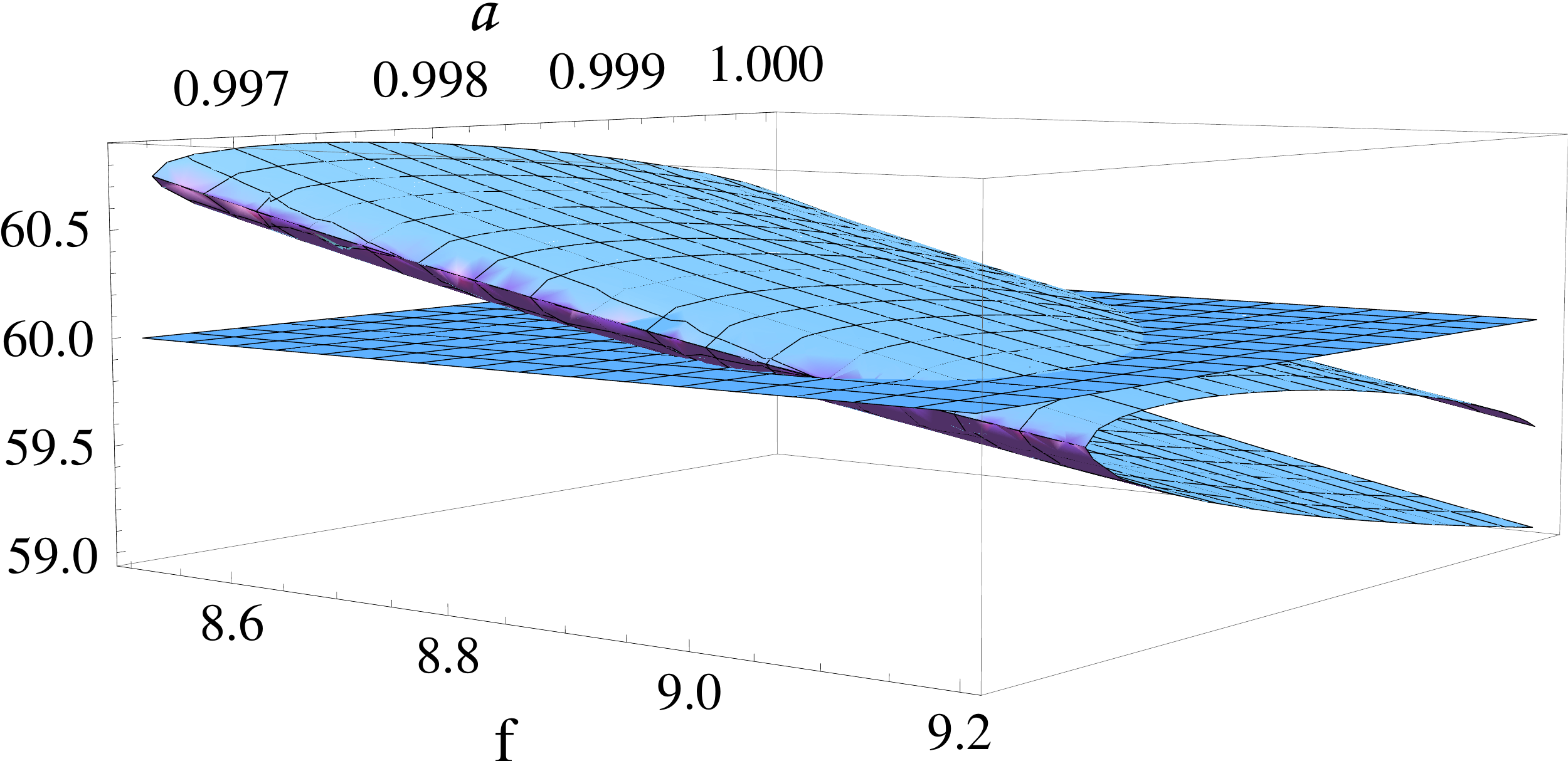}
        \caption{Plot of the number of e-folds Eq.\,(\ref{Ni}) as a function of the model parameters $f$ and $a$. The lower sheet corresponds to $N_1$ while the upper to $N_2$, the plane corresponds to $N=60$. We see that there is no important difference between them: Thus, in our discussion following Eq.\,(\ref{Ni})  we write the number of e-folds simply as $N$.}
\label{folded}
 \end{center}
\end{figure}

We now attempt a similar analysis as in Section\,\ref{NI}. The slow-roll parameters are now given by 
\begin{eqnarray}
\epsilon &=&\frac{1}{2}\left(\frac{a}{f}\right)^2\frac{1-c_{\mathrm{\phi}} ^{2}}{\left( 1+a\, c_{\mathrm{\phi}} \right)^2} ,
\label{HNIeps}%
\\
\eta &=&-\left( \frac{a}{f^2}\right)\, \frac{c_{\mathrm{\phi}}}{1+a c_{\mathrm{\phi}}} , 
\label{HNIeta}
\\
\xi_2 &=&-\left( \frac{a}{f^2}\right)^2\,\frac{1-c_{\mathrm{\phi}} ^{2}}{\left( 1+a\, c_{\mathrm{\phi}} \right)^2} ,\\
\xi_3 &=& +\left( \frac{a}{f^2}\right)^3\,\frac{1-c_{\mathrm{\phi}} ^{2}}{\left(1+a c_{\mathrm{\phi}}\right)^3} c_{\mathrm{\phi}} .
\end{eqnarray}%
From the expression for the scalar spectral index  Eq.\,(\ref{Ins})  we now get
\begin{equation}
\delta_{n_s}=\frac{a}{f^{2}}\, \frac{2 c_H+a(3-c_H^2)}{(1+ac_{H})^{2}}.  \label{spectral2}
\end{equation}
At $\phi _{H}$  Eq.\,(\ref{spectral2}) can be solved for $c_H$ \cite{Ross:2010fg} obtaining the following 
\begin{equation}
c_{1H}\equiv\frac{1-f^2\delta_{n_s}+\sqrt{1+3a^2-3(1-a^2)f^2\delta_{n_s}}}{a(1+f^2\delta_{n_s})},\text{\ }a\geqslant \frac{1}{3}  ,
\label{solution1}
\end{equation}
and 
\begin{equation}
c_{2H}\equiv\frac{1-f^2\delta_{n_s}-\sqrt{1+3a^2-3(1-a^2)f^2\delta_{n_s}}}{a(1+f^2\delta_{n_s})},\; a<1.  
\label{solution2}
\end{equation}
The first solution in the limit $a=1$ corresponds to NI. We thus study here the solution $c_{1H}$ only and leave $c_{2H}$ for the following sections. From the constraints  $-1<c_{1H}<1$ we get restrictions on the $f$ and $a$ parameters
\begin{eqnarray}
\frac{1}{\sqrt{2 \delta_{n_s}}} & < f < &\frac{1}{\sqrt{\delta_{n_s}}}\,,\quad\quad\quad \frac{1}{\sqrt{3}}\left(\frac{3f^2\delta_{n_s}-1} {f^2\delta_{n_s}+1}\right)^{1/2}< a \leq \frac{f^2\delta_{n_s}} {2-f^2\delta_{n_s}} ,
\label{c1Ha}%
\\
\frac{1}{\sqrt{\delta_{n_s}}} & < f\,,  &\quad\quad\quad\quad\quad\quad            \frac{1}{\sqrt{3}}\left(\frac{3f^2\delta_{n_s}-1} {f^2\delta_{n_s}+1}\right)^{1/2}< a \leq 1,
\label{c1Hb}
\end{eqnarray}%
or using $\delta_{n_s}\approx 0.035$
\begin{eqnarray}
3.78 & < f < &5.35\,,\quad\quad\quad \frac{1}{3}< a \leq 1,
\label{nc1Ha}%
\\
5.35 & < f\,,  &\quad\quad\quad\quad\quad\quad            \frac{1}{\sqrt{3}} < a \leq 1.
\label{nc1Hb}
\end{eqnarray}%
Thus, we see that the extension of NI corresponds to Eq.\,(\ref{solution1}) together with Eq.\,(\ref{c1Hb}). In NI the end of inflation is given by $\epsilon=1$, here from Eq.\,(\ref{HNIeps}) we find
\begin{eqnarray}
c_{e1} & = &-\frac{2f^2-\sqrt{   a^2-2(1-a^2)f^2 }}{a(1+2f^2)},
\label{ce1}%
\\
c_{e2} & = &-\frac{2f^2+\sqrt{   a^2-2(1-a^2)f^2 }}{a(1+2f^2)}.
\label{ce2}
\end{eqnarray}%
Thus, there are two values of $\phi_e$ which can meet the condition $\epsilon=1$ and end inflation.
The number of e-folds from $\phi_{{\rm H}}$ to the end of inflation at $\phi _{\mathrm{e}}$ is
\begin{equation}
N\equiv -\int_{\phi _H}^{\phi_e}\frac{V({\phi })}{V^{\prime }({\phi })}{d}{\phi }=\frac{f^2}{2a}\left((1+a)\ln \left( \frac{1-c_e}{1-c_H}\right)+(1-a)\ln \left( \frac{1+c_H}{1+c_e}\right)\right),
\label{N}
\end{equation}
where $c_{e}\equiv \cos \left( \frac{\phi _{e}}{f}\right)$. In ENI we thus have
\begin{equation}
N_i=\frac{f^2}{2a}\left((1+a)\ln \left( \frac{1-c_{ei}}{1-c_{1H}}\right)+(1-a)\ln \left( \frac{1+c_{1H}}{1+c_{ei}}\right)\right),\quad\quad i=1,2.
\label{Ni}
\end{equation}
The function $N_i(f,a)$ defines a two-dimensional sheet tightly folded (see Fig.\,\ref{folded})  as a consequence the number of e-folds $N_1$ is not very different from $N_2$. Thus in what follows we simply talk about the number of e-folds $N$ meaning any of them. In any case the requirement of $N=60$ e-folds of inflation begs for the parameter $a$ to be very close to 1. This is so because the ENI potential esentially differs from NI by a constant term. The inflationary epoch is controled by the derivative of the potential. Thus, a constant term in the potential is not very relevant in the determination of the number of e-folds and hence of the parameters $f$ and $a$; these are very close to the NI values. Not contributing to the cosmological constant problem leads us to impose $a=1$. Thus, a modification of NI to avoid the fine tuning problem requires a more drastic solution. In the following sections we continue our discussion in the context of HNI  \cite{Ross:2016hyb}, \cite{Ross:2009hg}, \cite{Ross:2010fg} where the end of inflation is not due to the saturation of the condition $\epsilon = 1$ but to the destabilisation of the inflaton direction by an extra{ \it waterfall} field. This approach liberates the inflationary sector from also ending inflation.
%%%%%%%%%%%%%%%%%%%%%%%%%%%%%%%%%%%%%%%%%%%
%%%%%%%%%%%%%%%%%%%%%%%%%%%%%%%%%%%%%%%%%%%

\section{Hybrid Natural Inflation, bounds in the restricted case} \label{restricted}

The first solution Eq.\,(\ref{solution1}) in the limit $a=1$ corresponds to \textit{Natural Inflation}. However to avoid the possibility of large gravitational corrections to the potential we will concentrate on the case $f<1.$\footnote{Recall that $f>\frac{1}{\sqrt{\delta_{n_s}}} \approx 5.3$ in NI.}  Thus, the relevant solution is the second one denoted by $c_{2H}$, hereafter $c_H$, given by 
\begin{equation}
c_H=\frac{1-f^2\delta_{n_s}-\sqrt{1+3a^2-3(1-a^2)f^2\delta_{n_s}}}{a(1+f^2\delta_{n_s})},
\label{ch}
\end{equation}
corresponds to $c_{2H}$ of Eq.\,(\ref{solution2}).
From Eq.\,(\ref{ch}) note that $c_H=1$ when $a=\frac{f^2\delta_{n_s}}{2-f^2\delta_{n_s}}$. Thus, in principle one can have very small $r$ (see Eq.\,(\ref{HNIeps})) and very low-scale of inflation (low $\Delta$, see Eq.\,(\ref{IA})) when $a$ gets close to $\frac{f^2\delta_{n_s}}{2-f^2\delta_{n_s}}$ for any value of $f<1$.  

The density perturbation for the HNI potential can be written as
\begin{equation}
\mathcal{P}_s(k) =\frac{1}{24\pi ^{2}}\frac{V(\phi)}{\epsilon(\phi)}=\frac{f^2 V_0\left(1+a\cos(\frac{\phi}{f})\right)^3}{12\pi^2 a^2\left(1-\cos(\frac{\phi}{f})^2\right)},
\label{contraste}
\end{equation}
which presents a minimum (see Fig.\,\ref{Espectro}) for $\phi_{min}$ given by
\begin{equation}
c_{{min}}=\cos\left(\frac{\phi_{min}}{f}\right)=\frac{1-\sqrt{1+3a^2}}{a} \approx -\frac{3}{2}a.
\label{cmin}
\end{equation}
This together with some other properties related with the non-monotonicity of the tensor-to-scalar ratio are studied in \cite{German:2015qjq}. 
From the equation for the number of e-folds Eq.\,(\ref{N}) we provisionally take (in what we call the restricted case)  the end of the first $8<N_{min}<60$ e-folds of inflation as given by $c_{min}$, this will determine $a$ for a given $f$. The formula for $N_{min}$ e-folds adapted from Eq.\,(\ref{N}) is
\begin{equation}
N_{min}=\frac{f^2}{2a}\left((1+a)\ln \left( \frac{1-c_{min}}{1-c_{H}}\right)+(1-a)\ln \left( \frac{1+c_H}{1+c_{min}}\right)\right).
\label{Nmin}
\end{equation}
We first study this formula analytically  for small $a$. The expansion of $c_H$ for small $a$ is given by
\begin{equation}
c_H= \frac{1}{2}\left(\frac{f^2 \delta_{n_s}}{a}\right)-\frac{3}{2}\left(1-\frac{5}{12}\left(\frac{f^2 \delta_{n_s}}{a}\right)^2\right)a+\cdot\cdot\cdot \approx \frac{1}{2}\left(\frac{f^2 \delta_{n_s}}{a}\right) +\mathcal{O}(a).
\label{chapp}
\end{equation}
For small $a$ the term $c_{min}$ is also small and negligible inside the $\log$ of Eq.\,(\ref{Nmin}) thus, $N_{min}$ can be approximated by
\begin{equation}
N_{min}\approx\frac{f^2}{2a}\ln \left( \frac{1+c_H}{1-c_{H}}\right)\approx \frac{1}{2 \delta_{n_s}}\left(\frac{f^2 \delta_{n_s}}{a}\right)^2+\frac{1}{24 \delta_{n_s}}\left(\frac{f^2 \delta_{n_s}}{a}\right)^4+\cdot\cdot\cdot,
\label{N8app}
\end{equation}
from where it follows that  $\frac{f^2}{a}$ is approximately constant 
\begin{equation}
\frac{f^2}{a}\approx    \frac{ \sqrt{6}}{\delta_{n_s}}  \left(N_s-1\right)^{1/2}.    
\label{constant}
\end{equation}
Here we have defined $N_s\equiv \sqrt{ 1+\frac{2}{3}N_{min}\delta_{n_s} }$ to simplify notation.
From
\begin{equation}
\Delta=\left(24\pi ^{2}\epsilon(\phi_H)\mathcal{P}_s(k_H)\right)^{1/4} =\left(12\pi^2 \mathcal{P}_s(k_H)  \left(\frac{a}{f}\right)^2  \frac{1-c_H^2  }{\left(1+a\,c_H\right)^2}\right)^{1/4},
\label{Delta}
\end{equation}
we get
\begin{equation}
\Delta\approx\left(\pi^2 \delta_{n_s}^2\mathcal{P}_s(k_H)\left(\frac{5-3N_s } {N_s -1}\right)+\mathcal{O}(a) \right)^{1/4} f^{1/2}.
\label{Deltaap}
\end{equation}
The behavior $\Delta \sim f^{1/2}$  means that $\Delta$ decreases more slowly than $f$. Thus, there is a point where $\Delta=f$ signaling the minimum value of $f$ consistent with $\Delta< f$. Solving $\Delta=f$ to lowest order in $a$
\begin{equation}
f_{min}\approx          \pi \delta_{n_s}\mathcal{P}^{1/2}_s(k_H)\left(\frac{5-3N_s}     {N_s-1}\right)^{1/2},
\label{fmin}
\end{equation}  
from where it follows that $N_{min}<76$, sufficient for our purpose. From Eqs.\,(\ref{constant}) and (\ref{Deltaap})  with $f=f_{min}$ we get the lower limit for $\Delta$ while setting $f=1/\pi$ (for consistency with $\Delta\phi<1$) gives the upper bound
\begin{equation}
\Delta_{min}\equiv \pi\delta_{n_s}\,\mathcal{P}^{1/2}_s(k_H)\left(\frac{5-3N_s}     {N_s-1}\right)^{1/2}  <\Delta< \left(\delta_{n_s}^2\mathcal{P}_s(k_H)\left(\frac{5-3N_s } {N_s -1}\right) \right)^{1/4}\equiv \Delta_{max}.
\label{Deltabounded}
\end{equation}
\begin{figure}[t!]
 \begin{center}
   \includegraphics[ width=8cm, height=5cm]{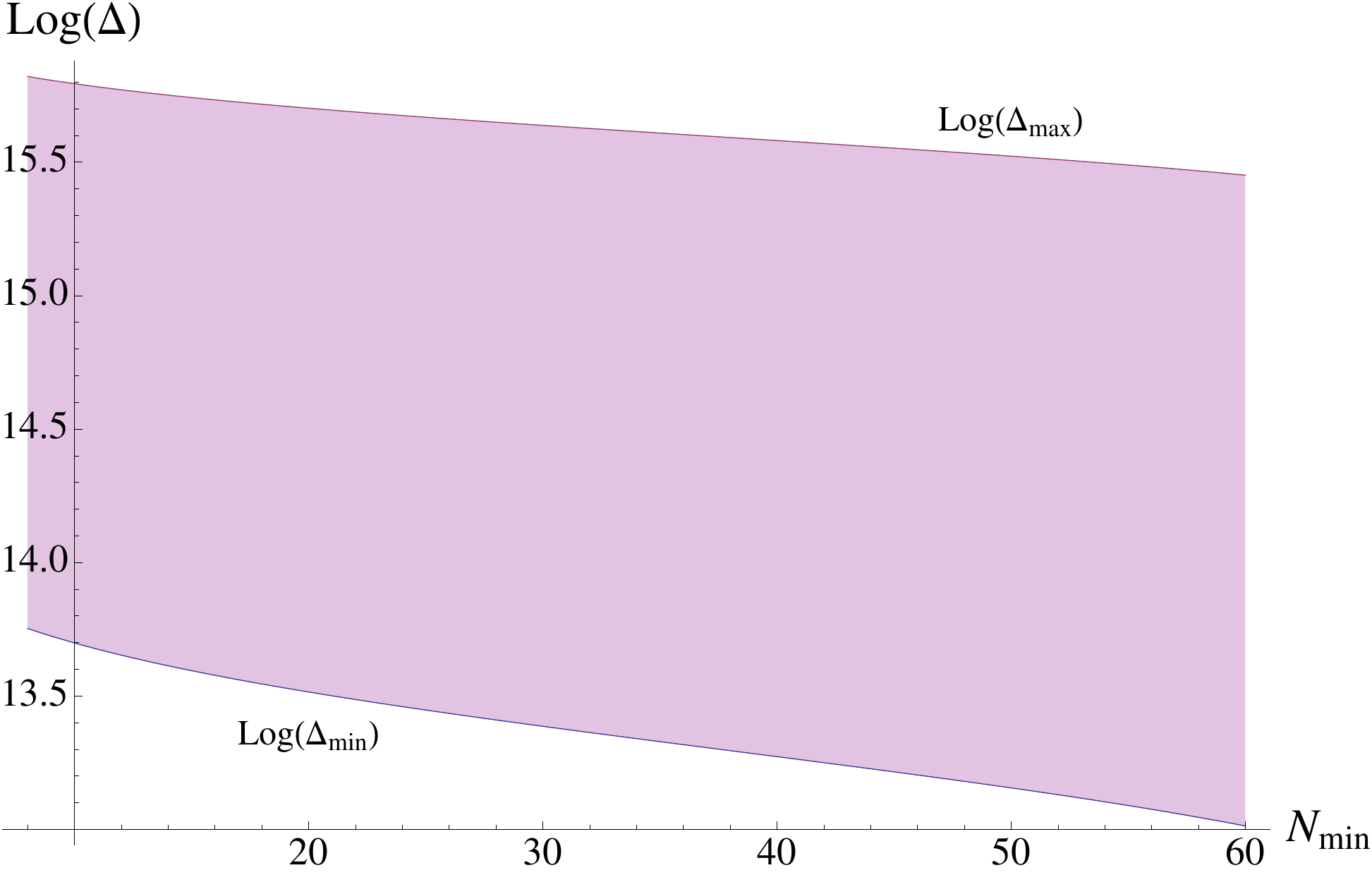}
    \includegraphics[ width=8cm, height=5cm]{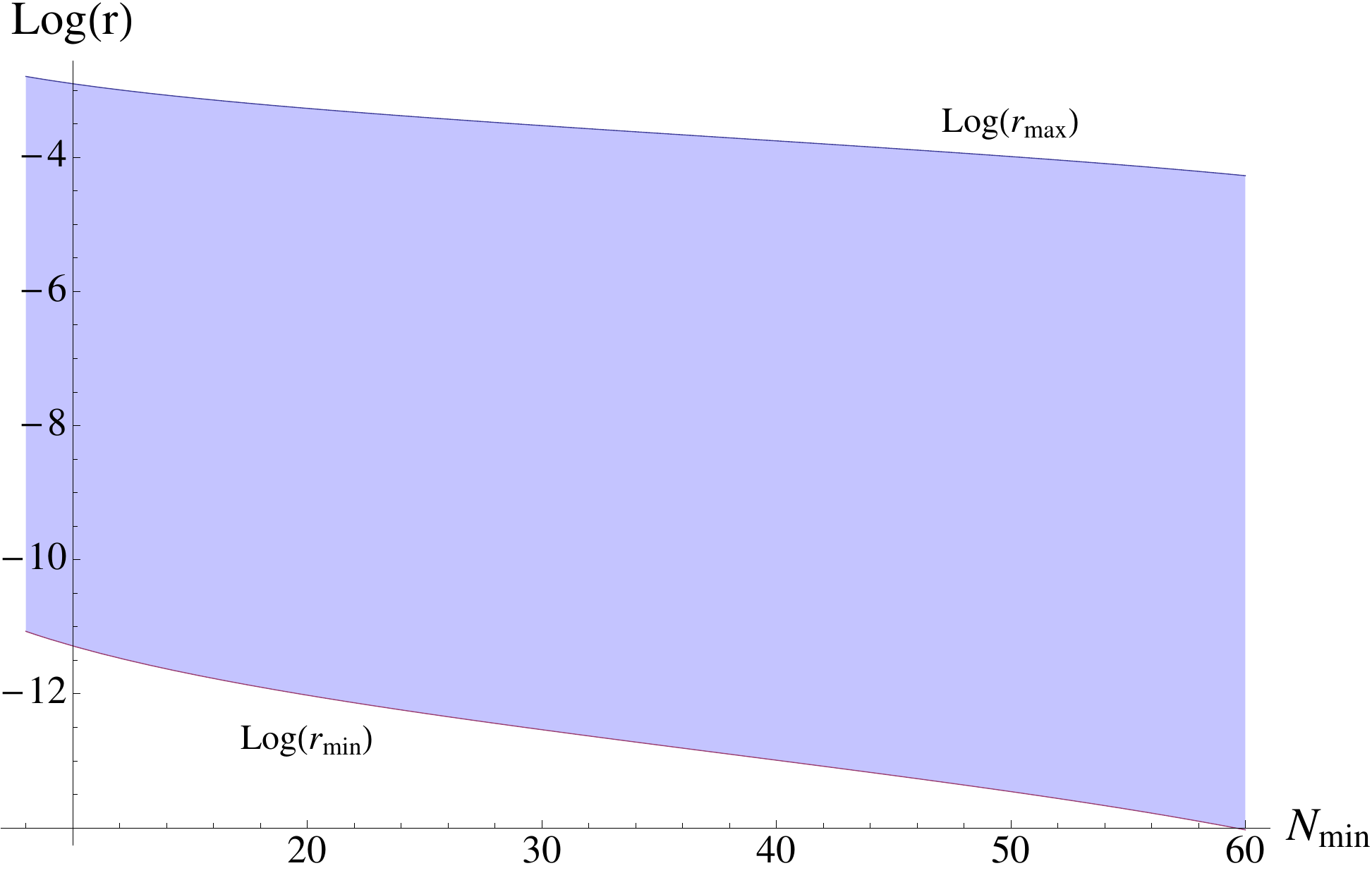}
        \caption{The logarithms of Eqs.\,(\ref{Deltabounded}) and (\ref{rbounded}) are shown as functions of $N_{min}$, where $8<N_{min}<60$ is the number of e-folds with decreasing spectrum counted from $\phi_H$ up to $\phi_{min} $.  }
\label{DB}
 \end{center}
\end{figure}
The upper limit follows simply from the requirement that $\Delta\phi<1$ and it is not derived from any stronger condition.
When there is a fixed number of e-folds $N_{min}<60$ from $\phi_H$ to the minimum at $\phi_{min}$ the scale of inflation as a function of $N_{min}$ is bounded as $\Delta_{min} < \Delta < \Delta_{max}$, see Fig.\,\ref{DB}a.
For small $a$ we can also see that the tensor-to-scalar ratio scales with $a$ as follows
\begin{equation}
r\approx \frac{2\sqrt{6}}{3}\,\delta_{n_s}\left(N_s -1\right)^{1/2}  \left(\frac{5-3N_s}{N_s-1}\right)a+ \mathcal{O}(a^2),
\label{rapp}
\end{equation}
and so becomes small for small $a$. From Eqs.\,(\ref{IA}) and (\ref{Deltabounded}) $r$ is bounded as follows, (see Fig.\,\ref{DB}b),
\begin{equation}
r_{min}\equiv \frac{2}{3}\pi^2\,\delta_{n_s}^4\,\mathcal{P}_s(k_H)\left(\frac{5-3N_s}{N_s-1}\right)^2 < r < \frac{2}{3\pi^2}\delta_{n_s}^2\left(\frac{5-3N_s}{N_s-1}\right)\equiv r_{max}.    
\label{rbounded}
\end{equation}
From Eq.\,(\ref{IA}) we see that the scale of inflation decreases much more slowly than the tensor-to-scalar ratio, $\Delta\equiv V_H^{1/4} \sim a^{1/4}$. 

In Ref.\,\cite{Hebecker:2013zda},  HNI was already given a detailed analysis for its feasibility to reach large values of $r$, for sub-Planckian axion decay constant and sub-Planckian field range. There, a constraint on $f$ coming from an embedding of the effective potential of 
Eq.~\eqref{pot} into a string theory  guide the authors to choose the fiducial bound $f\lesssim \frac{\sqrt{3}}{4\pi}$ giving an upper bound $r\simeq 7.6 \times 10^{-4}$  \cite{Hebecker:2013zda}. Using the same bound for $f$ our upper bound on $r$ changes by a factor of 3/16, from $r\simeq 1.6 \times 10^{-3}$ to $r\simeq 3 \times 10^{-4}$, for $N_{min}=8$ and $n_s=0.965$ as can be seen from Eq.~\eqref{rbounded}.

The expressions for the spectral indices are
\begin{eqnarray}
n_{sk} &\approx&\frac{\delta_{n_s}^2}{6} \left(\frac{5-3N_s}     {N_s-1}\right)+ \mathcal{O}(a), \\
n_{skk} &\approx&\frac{\delta_{n_s}^3}{6}  \left(\frac{5-3N_s}     {N_s-1}\right)+ \mathcal{O}(a), \label{nskkapp}
\end{eqnarray}
both are practically constant for small $a$. For numerical values see at the end of Section \ref{general}.
%%%%%%%%%%%%%%%%%%%%%%%%%%%%%%%%%%%%%%%%%%%
%%%%%%%%%%%%%%%%%%%%%%%%%%%%%%%%%%%%%%%%%%%

\section{Hybrid Natural Inflation, bounds in the general case} \label{general} 
\begin{figure}[t!]
 \begin{center}
   \includegraphics[ width=8cm, height=6cm]{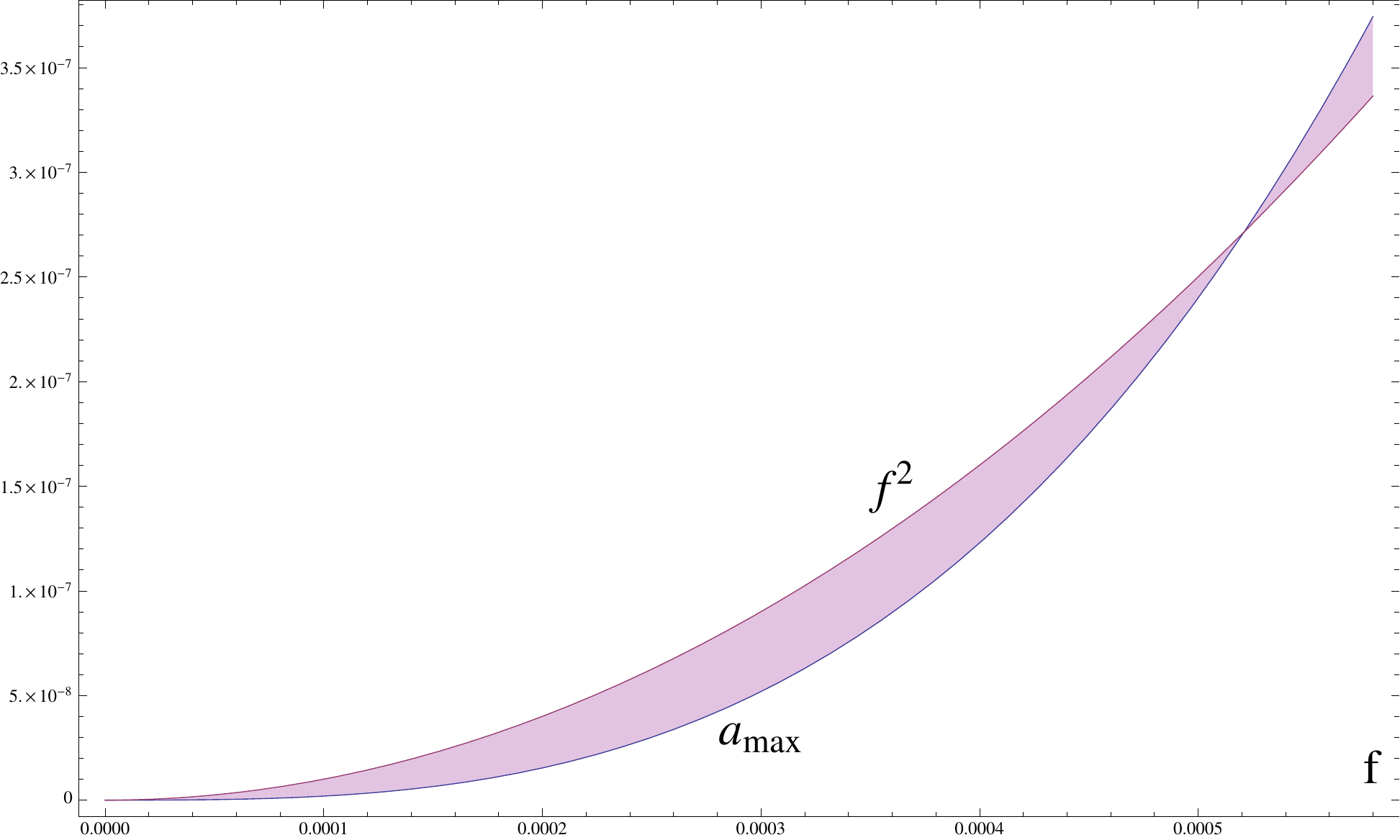}
    \includegraphics[ width=8cm, height=6cm]{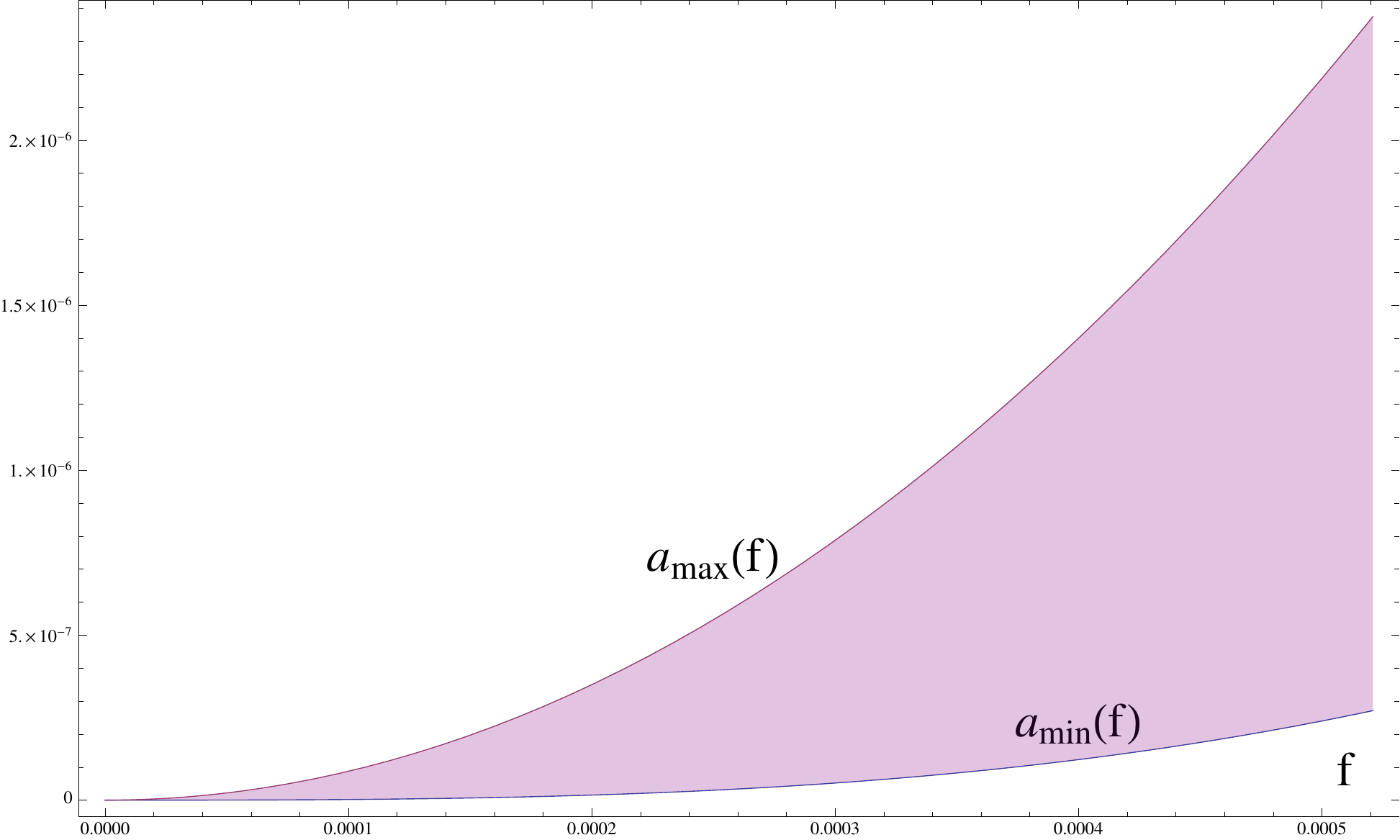}
        \caption{The figure on the left panel shows $a_{max}(f)$ as given by Eq.\,(\ref{amax}) and $f^2(f)$. The intersection point corresponds to the value of $f$ for which the parenthesis in the r.h.s. of Eq.\,(\ref{amax}) equals one,  this value is given by $f_{1}\approx 2\pi\sqrt{3 \mathcal{P}_s(k_H)}\approx 5.21 \times 10^{-4}$. At this point the slow-roll condition coming from $\eta<1$ is violated. For any value of $f<f_1$ the parameter $a$ can have values between $a_{min}$ and $a_{max}$ according to Eq.\,(\ref{abound1}), this is shown in the right hand panel above.}
\label{B2}
 \end{center}
\end{figure}
In the restricted case discussed in section \ref{restricted} the number of e-folds $8<N_{min}<60$ is counted from $\phi_H$ to $\phi_{min}$ with the remaining e-folds occurring with increasing spectrum. If all $N=60$ e-folds occur before $\phi$ reaches $\phi_{min}$ we can not use $\phi_{min}$ to count the number of e-folds and we are in the general case where the end of inflation is dictated by the waterfall sector of the theory.
From Eq.\,(\ref{Delta}), the equation $\Delta=f$  can actually be solved exactly for $a$ as a function of $f$, we denote this solution by $a_{max}$
\begin{equation}
a_{max}=f^2\left(\frac{\sqrt{3f^8+16f^2\pi^2 \mathcal{P}_s(k_H)-24 f^4\pi^2 \mathcal{P}_s(k_H)\delta_{n_s}  +48\pi^4\mathcal{P}_s(k_H)^2\delta_{n_s}^2}}{\sqrt{3}\left(f^6+8\pi^2 \mathcal{P}_s(k_H)-4 f^2\pi^2 \mathcal{P}_s(k_H)\delta_{n_s}\right)} \right),
\label{amax}
\end{equation}
but contrary to the $N_{min}$ case $a/f^2$ is not approximately constant but clearly depends on $f$. The condition $\Delta < f$ restricts the value of the parameter $a$ to be less than $a_{max}$ for a given $f$.  On the other hand the slow-roll condition $a/f^2<1$ (coming from $\eta<1$) restricts $f$ such that the formula for $a_{max}$ is only valid up to the value of $f$, denoted by $f_1$, such that the term in parenthesis in the r.h.s. of Eq.\,(\ref{amax}) equals one, this occurs for $f_{1}\approx 2\pi\sqrt{3 \mathcal{P}_s(k_H)}\approx 5.21 \times 10^{-4}$ (left panel in Fig.\,\ref {B2}). Thus $a_{max}$ is well approximated by
\begin{equation}
a_{max}\approx    \frac{f^2\delta_{n_s}}{2} \sqrt{1+\frac{f^2}{3\pi^2\delta^2_{n_s}\,\mathcal{P}_s(k_H)}}, \quad\quad f < f_1.
\label{amaxapp}
\end{equation}
For $f$ larger than $f_{1}$  the condition $\Delta < f$ is always satisfied whenever $a$ is restricted by the stronger condition $a<f^2$. 
Thus for a given $f < f_{1}$  we get the bounds for $a$
\begin{equation}
a_{min}\equiv \frac{f^2\delta_{n_s}}{2-f^2\delta_{n_s}} < a < a_{max}\,, \quad\quad f < f_1,
\label{abound1}
\end{equation}
while for $f > f_{1}$ the bounds are 
\begin{equation}
\frac{f^2\delta_{n_s}}{2-f^2\delta_{n_s}} < a < f^2\,, \quad\quad f > f_1.
\label{abound2}
\end{equation}
the lower bound in Eqs.\,(\ref{abound1}) and (\ref{abound2}) comes from requiring $c_H < 1$ in  Eq.\,(\ref{ch}). Thus as we lower the value of $f$ the range of possible $a$ values reduces according to the bounds above. In Fig.\,\ref {B2} (right panel) for any value of $f$ all possible $a$-values define a vertical line in the shaded region.
\begin{figure}[t!]
 \begin{center}
   \includegraphics[ width=8cm, height=6cm]{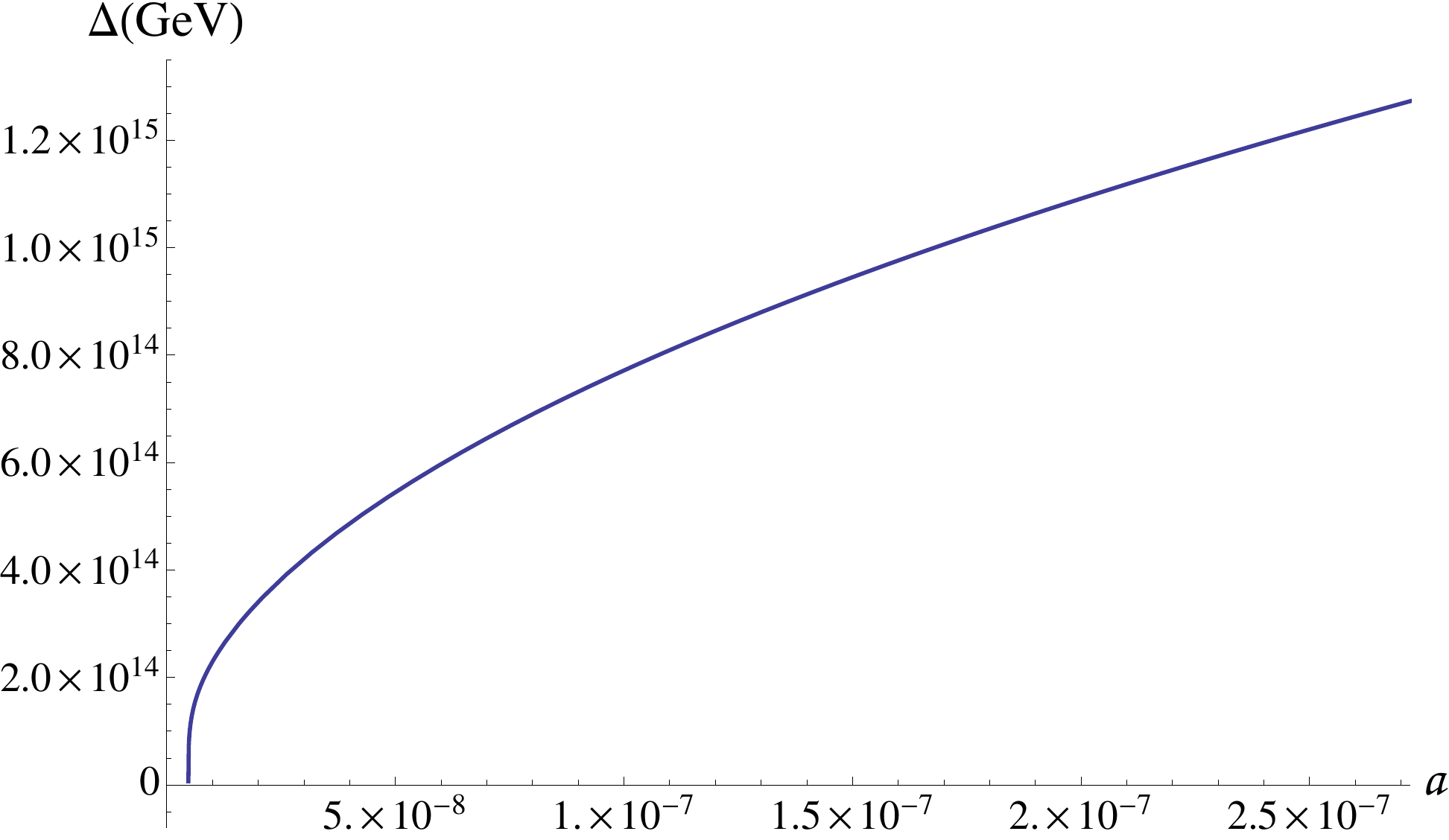}
        \caption{For a given value of $f$ (in this case $f=f_1\approx 5.21 \times 10^{-4}$) the scale of inflation $\Delta$ sweeps all range of values, from vanishingly small up to the $GUT$ scale, as $a$ goes from  $a_{min}$ to $a_{max}$ according to Eq.\,(\ref{abound1}).}
\label{B3}
 \end{center}
\end{figure}
As a consequence $\Delta$ will be bounded by $f(<f_1)$ whenever $a<a_{max}$ , in this case $r$ is bounded as follows
\begin{equation}
0 < r < \frac{2 f^4}{3\pi^2\, \mathcal{P}_s(k_H)}\,, \quad\quad\quad\quad f < f_1,
\label{rGeneralBounded}
\end{equation}
while for $f > f_{1}$ the scale of inflation can have any value in  the interval $0<\Delta <f $ whenever $a<f^2$. 
Thus, in the general case $\Delta$ is unbounded from below since this was a consequence of fixing the number of e-folds from $\phi_H$ to $\phi_{min}$ to a certain number $N_{min}\leq 60$. In the general case  the end of inflation can occur before $\phi$ reaches the minimum of the spectrum at $\phi_{min}$ and no relation beetwen $a$ and $f$ for a fixed $N$ can be found because $\phi_e$ is undetermined. The scale of inflation $\Delta$ is vanishingly small for $a$ approaching $\frac{f^2\delta_{n_s}}{2-f^2\delta_{n_s}}$ because $c_H$ tends to one in that limit. As $a$ goes from $ \frac{f^2\delta_{n_s}}{2-f^2\delta_{n_s}}$ to its upper limit $c_H$ diminishes from 1 and the scale $\Delta$ grows from very small values, this is how the potential is able to cover the whole range of inflationary scales (see Fig.\,\ref {B3}).

Finally the reader would have noticed that we can apply Eq.\,(\ref{amax}) directly to the restricted case, substituting in Eq.\,(\ref{Nmin}) for the number of e-folds from $\phi_H$ up to $\phi_{min}$, extract the values of $f$ which accommodate from 60 to 8 e-folds and evaluate all other quantities of interest. While this is certainly possible and the numerical ranges are given below in Section \ref{restricted} we wanted to obtain an approximated {\it analytical} expression which can teach us more than a few simple numbers. The corresponding ranges for the {\it lower} bounds when the number of e-folds $N_{min}$ go from 8 to 60 are given below (see also Fig.\, \ref{B2}),
\begin{eqnarray}
 5.75 \times 10^{-6}&< f < & 2.32\times 10^{-5}, \\
  6.84\times 10^{-13}&< a_{max} < &2.58\times 10^{-11}, \\
   60&>N_{min}>&8, \\
1.40  \times 10^{13}GeV&<\Delta<&5.67\times 10^{13}GeV, \\
3.22    \times 10^{-14}&<r<&8.57 \times 10^{-12}, \\
2.43      \times 10^{-4}&<n_{sk}<&3.97\times 10^{-3}, \\
8.52        \times 10^{-6}&<n_{skk}<&1.39\times 10^{-4},  \label{ranges}
\end{eqnarray}
which compare very well with values obtained from the analytical approximations of Section\,\ref{restricted}. Clearly, in the general case we cannot do this because the formula for the number of e-folds Eq.\,(\ref{N}) involves $c_{e}$ which can depend on parameters different from $a$ and $f$.

%%%%%%%%%%%%%%%%%%%%%%%%%%%%%%%%%%%%%%%%%%%
%%%%%%%%%%%%%%%%%%%%%%%%%%%%%%%%%%%%%%%%%%%

\section{PBH constraints and low scales of inflation} \label{PBH} 

The steepening of the scalar spectrum in HNI (Fig.\,\ref{Espectro}) gives rise to a positive running allowing for the possibility of primordial black hole production during inflation   \cite{Kohri:2007qn}, \cite{Ross:2016hyb}.  In terms of the wave number $k$  the scalar power spectrum at first order in the SR parameters is given by
\begin{equation}
\mathcal{P}_s(k)=A_s\left( \frac{k}{k_H}\right)^{(n_s-1) + \frac{1}{2}n_{sk}  \ln\left(\frac{k}{k_H}\right) +\, \cdot \, \cdot \, \cdot  }.\label{power}
\end{equation}
Due to the constraint coming from the possible over-production of primordial black holes (PBHs) at the end of inflation the Taylor expansion of the power spectrum around its  value at horizon crossing, $N_{H}\approx 60$ is bounded by \footnote{To lowest order in slow-roll $d/dN =- d / d\ln k$. The next order term in the expansion of Eq.~\eqref{ps:expansion}, involving the parameter $n_{skk}$, is subdominant.},
\begin{equation}
\label{ps:expansion}
C_{PBH}\equiv \ln \left[\frac{\mathcal{P}_s(0)}{\mathcal{P}_s(N_H)}\right] = (n_s- 1) N_H + \frac{1}{2} n_{sk} N_H^2\leq 14,
\end{equation}
where $\mathcal{P}_s(N= 0)\simeq 10^{-3}$ (see also Refs.~\cite{Josan:2009qn,Carr:2009jm}) evolves from the initial value $\mathcal{P}_s(N_H) \approx 10^{-9}$. 
This gives the bound $n_{sk}< 10^{-2}$. For the HNI potential this constraint can be written as
\begin{equation}
C_{PBH}=\ln\left[\frac{(1-c_H^2)(1+a\, c_e)^3}{(1-c_e^2)(1+a\, c_H)^3} \right],
\label{pbh}
\end{equation}
and can be easily satisfied for all  cases discussed since here $C_{PBH}<3$. A more stringent bound may be set by PBHs produced during reheating \cite{Hidalgo:2017dfp}, \cite{Carr:2017edp}. However, this depends on the specific reheating model and its associated equation of state. Such restrictions will be explored elsewhere.

Low scales of inflation $\Delta$ can be obtained when $r$ is very small since, from Eq.\,(\ref{IA}), $\Delta\sim r^{1/4}$. On the other hand, from Eq.\,(\ref{HNIeps}) we see that $r$ is small when $c_H$ is very close to 1. This occurs for $a$ approaching the lower bound in Eqs.\,(\ref{abound1}) and (\ref{abound2}). Notice that the parameters $a$ and the scale of symmetry breaking $f$ need not be very small to give small inflationary scales, instead they should be closely related by $a\approx \frac{f^2 \delta_{ns}}{2-f^2 \delta_{ns}}$. In the absence of a mechanism which sets $a$ close to $\frac{f^2 \delta_{ns}}{2-f^2 \delta_{ns}}$ this is fine tuning which is equivalent to starting inflation with $\phi_H$ very close to the origin.
In any case the value of $\phi_{H}$ should exceed the quantum fluctuations of the field $\delta\phi\approx\frac{H}{2\pi}\approx\frac{\Delta^2}{2\pi\sqrt{3}}$. From Eqs.\,(\ref{HNIeps}) and (\ref{IA}) we get the small $\phi$ behaviour 
\begin{equation}
r \approx \frac{\delta_{ns}^2}{2} \phi_{H}^2=\frac{2}{3 \pi^2A_s}\Delta^4,
\label{rsmall}
\end{equation}
from where it follows that
\begin{equation}
\phi_{H}\approx\left(\frac{4}{3 \pi^2A_s \delta_{ns}^2}\right)^{1/2}\Delta^2\approx 2.2\times10^5\Delta^2>>\frac{\Delta^2}{2\pi\sqrt{3}}=\delta\phi.
\label{rsmall}
\end{equation}

%%%%%%%%%%%%%%%%%%%%%%%%%%%%%%%%%%%%%%%%%%%
%%%%%%%%%%%%%%%%%%%%%%%%%%%%%%%%%%%%%%%%%%%

\section{Summary and conclusions} \label{conclusions} 

\noindent
An interesting characteristic of Hybrid Natural Inflation is that the tensor-to-scalar ratio is a non-monotonic function of $\phi$ with the parameter $\epsilon(\phi)$ developing a {\it maximum}. A consequence of this is that the scalar spectrum of density fluctuations develops a  {\it minimum} for some value $\phi_{min}$ of the inflaton. The value of $\phi_{min}$ is always slightly larger than the value of $\phi$ at the inflection point of the potential at $\phi_{I}=\pi/2$. Since the scalar spectrum has been observed to be decreasing during some 8 e-folds of observable inflation we can determine upper bounds for the scale of inflation and for the tensor-to-scalar ratio. In the {\it restricted} case considered in section \ref{restricted} we do this by requiring a minimum of $8<N_{min}<60$ e-folds of inflation from  the scale $\phi_{H}$, at which observable perturbations are produced, to $\phi_{min}$ where the spectrum stops decreasing. The remaining $0-52$ e-folds of inflation would occur with an steepening spectrum thus care is taken to not over-produce primordial black holes. The condition of having $N_{min}$ e-folds of inflation with a decreasing spectrum fixes the parameter $a$ once the symmetry breaking scale $f$ is determined. A minimum value for $f$ can be obtained by the requirement that the inflationary energy scale $\Delta$ is bounded by $f$. This allows to determine {\it lower} bounds  for the inflationary energy scale and the tensor-to-scalar ratio. 
The general case is discussed in section \ref{general} where we find upper bounds for $\Delta$ as well as $r$. In the general case the inflationary scale $\Delta$ is unbounded from below and can go all the way to vanishing values. The lower bound of section \ref{restricted} is a consequence of fixing the number of e-folds from $\phi_H$ to $\phi_{min}$ to a certain number $N_{min}\leq 60$. In the general case  the end of inflation can occur before $\phi$ reaches the minimum of the spectrum at $\phi_{min}$ and no relation beetwen $a$ and $f$ for a fixed $N$ can be found because $\phi_e$ is undetermined. The scale of inflation $\Delta$ is vanishingly small for $a$ approaching $\frac{f^2\delta_{n_s}}{2-f^2\delta_{n_s}}$ because $c_H$ tends to one in that limit. As $a$ goes from $ \frac{f^2\delta_{n_s}}{2-f^2\delta_{n_s}}$ to its upper limit $c_H$ diminishes from 1 and the scale $\Delta$ grows from very small values, this is how the potential is able to sweep the whole range of inflationary scales. By finding lower as well as upper bounds for the parameters $a$ and $f$ we can clearly understand how the scale of inflation in HNI is able to cover the complete range of values, from vanishingly small up to the GUT scale.

\section{Acknowledgements}

We are grateful to SNI for partial financial support. AHA also acknowledges a VIEP-BUAP-HEAA-EXC17-I research grant. We acknowledge financial support from PAPIIT-UNAM grant IA-103616 {\it Observables en cosmolog\'ia relativista} as well as CONACyT grants 269639 and 269652.

\end{document}